\documentclass[a4paper,12pt,leqno]{article}
\usepackage{theorem}
\usepackage{latexsym,amssymb,amsfonts,amsmath}
\usepackage[dvips]{graphicx}
\newtheorem{thm}{Theorem}[section]
\newtheorem{lem}[thm]{Lemma}
\newtheorem{cor}[thm]{Corollary}
\newtheorem{pro}[thm]{Proposition}

\theoremheaderfont{\scshape}
\newcommand{\RM}{\mathbb{R}}
\newcommand{\ZM}{\mathbb{Z}}

\newcommand{\CM}{\mathbb{C}}

\newcommand{\ket}[1]{|#1\rangle}

\title{{\Large {\bf Orthogonal polynomials induced by discrete-time quantum walks in one dimension}
}}
\author{
{\small Masatoshi HAMADA, Norio KONNO$^{\dagger}$\footnote{To whom correspondence should be addressed. E-mail: konno@ynu.ac.jp} } \\
{\scriptsize Department of Applied Mathematics, 
Faculty of Engineering, 
Yokohama National University}\\
{\scriptsize Hodogaya, Yokohama 240-8501, Japan} \\
{\scriptsize d06sb107@ynu.ac.jp, konno@ynu.ac.jp$^{\dagger}$}\\
{\small Wojciech M{\L}OTKOWSKI} \\
{\scriptsize Mathematical Institute, University of Wroc{\l}aw } \\
{\scriptsize Pl. Grunwaldzki 2/4, 50-384 Wroc{\l}aw, Poland} \\
{\scriptsize mlotkow@math.uni.wroc.pl}\\
} 

\date{\empty }
\begin{document}
\maketitle

\begin{small}
\par\noindent
{\bf Abstract}. In this paper we obtain some properties of orthogonal polynomials given by a weight function which is a limit density of a rescaled discrete-time quantum walk on the line. 
\end{small}
\begin{small}
\footnote[0]{
{\it Abbr. title:} Orthogonal polynomials induced by quantum walks}
\end{small}

\setcounter{equation}{0}
\section{Introduction}
A quantum walk is the quantum analog of a classical random walk. Quantum walks are expected to play an important role in the field of quantum algorithms. A number of benefits for such walks are already known. Reviews and books on quantum walks are Kempe \cite{Kempe2003}, Kendon \cite{Kendon2007}, Konno \cite{Konno2008a, Konno2008b}, Venegas-Andraca \cite{VAndraca2008}, for examples. There are two types of quantum walks. One is the discrete-time walk and the other is the continuous-time one. Here we focus on the case of a discrete-time walk on $\ZM$, where $\ZM$ is the set of integers. Ambainis {\it et al.} \cite{AmbainisEtAl2001} investigated the quantum walk intensively. In the present paper, we consider some properties of orthogonal polynomials given by a weight function which is a limit density of a rescaled discrete-time quantum walk on $\ZM$. Recently Cantero {\it et al.} \cite{CanteroEtAl2009} showed that the theory of matrix-valued orthogonal polynomials associated with a certain kind of unitary matrices, i.e., the CMV matrices, is a natural tool to study the discrete-time quantum walk on the line. It would be interesting to know a relation between their approach and our results.

The rest of the paper is organized as follows. In Sect. 2, we define the quantum walk and explain the weak limit theorem. Section 3 treats a symmetric qunatum walk. In Sect. 4, we give a result for an asymmetric case. Section 5 is devoted to a general case. 

\section{Quantum walk}
First we give a definition of one-dimensional discrete-time quantum walk. The time evolution of the quantum walk is defined by the following matrix: 
    \[
        U =  \begin{bmatrix}
                  a & b \\
                  c & d 
               \end{bmatrix}
        \in \text{U}(2),
    \]
where $a, b, c, d \in \mathbb{C}$ and $\text{U}(2)$ is the set of 2 $\times$ 2 unitary matrices. Here $\CM$ is the set of complex numbers. The unitarity of $U$ gives 
    \[	
        |a|^2 + |b|^2 = |c|^2 + |d|^2 = 1, \>\> a\bar{b} + c\bar{d} = 0, \>\>
        c = -\triangle \bar{b}, \>\> d = \triangle \bar{a},
    \]
where $\bar{z}$ is the complex conjugate of $z \in \mathbb{C}$ and $\triangle = ad - bc$. The quantum walk is a quantum version of the classical random walk with additional degree of freedom called chirality. The chirality takes values left and right, and it means the direction of the motion of the particle. At each time step, if the particle has the left chirality, it moves one step to the left, and if it has the right chirality, it moves one step to the right. Let define
\begin{eqnarray*}
\ket{L} = 
\left[
\begin{array}{cc}
1 \\
0  
\end{array}
\right],
\qquad
\ket{R} = 
\left[
\begin{array}{cc}
0 \\
1  
\end{array}
\right], 
\end{eqnarray*}
so $U$ acts on two chiralities as follows: 
\begin{eqnarray*}
U \ket{L} = a \ket{L} + c \ket{R}, \qquad U \ket{R} = b \ket{L} + d \ket{R},  
\end{eqnarray*}
where $L$ and $R$ refer to the right and left chirality state respectively. Here the set of initial qubit states is defined by 
\[
\Phi = \left\{ \varphi =
{}^T \left[\alpha, \beta \right] \in \CM^2 : |\alpha|^2 + |\beta|^2 =1 
\right\},
\]
where $T$ is the transposed operator. Let $X_n (= X_n^{\varphi})$ be the quantum walk at time $n$ starting from the origin with the initial state $\varphi \in \Phi$. To explain $X_n$ more precisely, we introduce $P$ and $Q$ given by 
    \[
        P = \begin{bmatrix}
                 a & b \\
                 0 & 0 \\
              \end{bmatrix}, \>\>
        Q = \begin{bmatrix}
                  0 & 0 \\
                  c & d \\
               \end{bmatrix},
    \]
with $U = P + Q$. The important point is that $P$ (resp. $Q$) represents that the particle moves to the left (resp. right). Let $\Xi_{n}(l, m)$ denote the sum of all paths in the trajectory consisting of $l$ steps left and $m$ steps right. In fact, for time $n = l+m$ and position $x=-l + m$, we have 
    \[
        \Xi_{n}(l, m) = \sum _{l_{j}, m_{j}}
         P^{l_{1}}Q^{m_{1}} P^{l_{2}}Q^{m_{2}} \dots P^{l_{n}}Q^{m_{n}},
    \]
summed over all non-negative integers $l_j$ and $m_j$ satisfying $l_1 + \cdots +l_n = l$ and $m_1 + \cdots +m_n = m$ with $l_j + m_j =1.$ The definition gives 
    \[
        \Xi_{n+1}(l, m) = P \> \Xi_{n}(l-1, m) + Q \> \Xi_{n}(l, m-1).
    \]
The probability that $X_{n} =x$ is defined by 
    \[
        P(X_{n} =x) = || \Xi_{n}(l, m)\varphi ||^2.
    \]
We should remark that there is a strong structural similarity between quantum walks and correlated random walks, see Konno \cite{Konno2009}. A typical example of the quantum walk is the Hadamard walk defined by the Hadamard gate $U=H$:
\begin{eqnarray*}
H=\frac{1}{\sqrt2}
\left[
\begin{array}{cc}
1 & 1 \\
1 &-1 
\end{array}
\right].
\end{eqnarray*}
The walk has been extensively investigated in the study of the quantum walk. 

Quantum walks behave quite differently from classical random walks. For example, in the classical case, the probability distribution is a binomial one. On the other hand, the probability distribution of the quantum walk has a complicated and oscillatory form. For the classical case, the well-known central limit theorem holds. For the quantum case, a corresponding weak limit theorem was shown by Konno \cite{Konno2002, Konno2005} in the following way: 

\begin{thm}
If $abcd \not= 0$, then 
\[
        \lim_{n \to \infty} P( u \leq X_{n} / n \leq v ) 
        = \int_{u}^{v} \> \{ 1 - c(a, b, \varphi) x\} \> k(x : |a|) \> dx,
\]
where 
\[
k (x : r) = \frac{\sqrt{1-r^2}}{ \pi ( 1-x^2 ) \sqrt{r^2-x^2} } \> I_{(-r, \> r)}(x), \qquad c(a,b : \varphi) = |\alpha|^2 - |\beta|^2 + \frac{a \alpha \overline{b \beta} + \overline{a \alpha}b \beta}{|a|^2}, 
\]
and $\varphi = {}^T[\alpha, \beta]$. Here $I_A (x) = 1 \> (x \in A), \> = 0 \> (x \not\in A).$ In other words, the limit density of $X_n/n$ is $f_{\infty} (x) \equiv \{ 1 - c(a, b, \varphi) x\} \> k(x : |a|).$
\end{thm}

\section{Symmetric case}
In this section we consider the symmetric case of the limit density
$f_{\infty} (x)$, i.e., $c(a,b\!:\!\varphi)=0$.
Let $\mu$ be the probability measure on the real line $\RM$,
with the density $k(x\!:\!r)$, where $0<r<1$.

\begin{thm}\label{thm1} Let $G_{\mu}(z):=\int_{\mathbb{R}}\frac{d\mu(x)}{z-x}$
be the Stielties transform of $\mu$. Then
\begin{gather*}
    G_{\mu}(z) = {z(z^2-r^2)-\sqrt{1-r^2}\sqrt{z^2-r^2}\over(z^2-1)(z^2-r^2)}.
\end{gather*}
Moreover, $G_{\mu}$ admits the following expansion as continued fraction:
\begin{gather*}
    G_{\mu}(z) = 
        \cfrac{1}{z-\cfrac{1-\sqrt{1-r^2}}{z - 
            \cfrac{\left(\sqrt{1-r^2}-1+r^2\right)/2}{   
                z-\cfrac{r^2/4}{z-\cfrac{r^2/4}{\ddots}}}}}.
\end{gather*}
\end{thm}

Before the proof we will derive some consequences.

\begin{cor}\label{cor2}
The monic orthogonal polynomials for $\mu$ are given by: $P_0 (x)=1,$ 
\begin{align*}
xP_{n}(x)=P_{n+1}(x)+\gamma_{n-1}P_{n-1}(x),
\end{align*}
where 
\begin{align*}
\gamma_0=1-\sqrt{1-r^2},\quad\gamma_1=\frac{\sqrt{1-r^2}(1 -\sqrt{1-r^2})}{2},
\quad \gamma_{n}=\frac{r^2}{4}\quad\hbox{for $n \ge 2$,}
\end{align*}
under convention that $P_{-1}(x)=0$ and $\gamma_{-1}=0$.
\end{cor}

In particular, for the Hadamard walk case $(r = 1/\sqrt{2})$ we have
\begin{align*}
\gamma_{0}=\frac{2-\sqrt{2}}{2},\quad\gamma_{1}=\frac{\sqrt{2}-1}{4},
\quad\gamma_{n}=\frac{1}{8}\quad\hbox{for $n \geq 2 $.}
\end{align*}
Then we can compute a few first orthogonal polynomials:
\begin{align*}
P_{0}(x) &= 1, \qquad P_{1}(x) = x, \\
P_{2}(x) &= x^2 + \frac{-2+\sqrt{2}}{2}, \\
P_{3}(x) &= x^3 + \frac{-3+\sqrt{2}}{2^2}x, \\
P_{4}(x) &= x^4 + \frac{-7+2\sqrt{2}}{2^3}x^2 + \frac{2-\sqrt{2}}{2^4}, \\
P_{5}(x) &= x^5 + \frac{-4+\sqrt{2}}{2^2}x^3 + \frac{7-3\sqrt{2}}{2^5}x, \\
P_{6}(x) &= x^6 + \frac{-9+2\sqrt{2}}{2^3}x^4 + \frac{21-8\sqrt{2}}{2^6}x^2 + \frac{-2+\sqrt{2}}{2^7}.
\end{align*}

From Corollary {\rmfamily \ref{cor2}}, we can easily compute the following generating function: 
\begin{align*}
Q(x,z)=\sum_{n=0}^{\infty} P_{n}(x)z^n.
\end{align*}

\begin{cor}
\begin{align*}
\left( 4 - 4xz + r^2 z^2 \right) Q(x,z)
= 4 + \left( r^2 -4 + 4\sqrt{1-r^2} \right) z^2 + \left( 2 - 2\sqrt{1-r^2} - r^2 \right) xz^3.
\end{align*}
In particular, for the Hadamard walk
\begin{align*} 
\left( 4 - 4xz + \frac{1}{2}z^2 \right) Q(x, z)
= 4 + \left( -\frac{7}{2} + 2\sqrt{2} \right) z^2 + \left( \frac{3}{2} - \frac{1}{\sqrt{2}} \right) xz^3. 
\end{align*}
\end{cor}

The following fact is useful for computing $G_{\mu}(z)$.

\begin{lem}
Define
\begin{align*}
    A(z)=\cfrac{1}{z-\cfrac{r^2/4}{z-\cfrac{r^2/4}{\ddots}}}.
\end{align*}
Then we have 
\begin{align*}
    A(z)=\frac{2z}{r^2}-\frac{2}{r^2}\sqrt{z^2-r^2}.
\end{align*}
\end{lem}

\noindent
\textit{Proof.} The definition of $A(z)$ yields
\begin{align*}
    A(z)={1\over z-r^2A(z)/4},
\end{align*}
so we have
\begin{align*}
    r^2A(z)^2-4zA(z)+4=0.
\end{align*}
Therefore
\begin{align*}
    A(z)=\frac{2z}{r^2}\pm\frac{2}{r^2}\sqrt{z^2-r^2},
\end{align*}
where the sign is choosen in such a way that $A$ maps 
the upper half plane into the lower half plane, see \cite{ak,do}.
\smallskip

\textit{Proof of Theorem {\rmfamily \ref{thm1}}.}
Denoting the continued fraction by $G(z)$ we have
\begin{align*}
G(z) 
&=\cfrac{1}{z-\cfrac{1-\sqrt{1-r^2}}{z-\cfrac{\left(\sqrt{1-r^2}-1+r^2\right)/2}{z-A(z)r^2/4}}}
\\
&=\cfrac{1}{z-\cfrac{1-\sqrt{1-r^2}}{z-\cfrac{\sqrt{1-r^2}-1+r^2}{z+\sqrt{z^2-r^2}}}}
\\
&=\cfrac{1}{z-\cfrac{1-\sqrt{1-r^2}}{z-\left(\sqrt{1-r^2}-1+r^2\right)\left(z-\sqrt{z^2-r^2}\right)/r^2}} 
\\
&=\cfrac{z+\sqrt{1-r^2}\sqrt{z^2-r^2}}{z^2-r^2+z\sqrt{1-r^2}\sqrt{z^2-r^2}} 
\\
&=\cfrac{z\left(z^2-r^2\right)-\sqrt{1-r^2}\sqrt{z^2-r^2}}{\left(z^2-1\right)\left(z^2-r^2\right)}, 
\end{align*}
which is equal to the right hand side of the first formula.
Now, using the standard technique (see \cite{ak,do}) one can
verify that $G(z)$ is Stielties transform for $\mu$,
which concludes the proof of Theorem {\rmfamily \ref{thm1}}.

\smallskip
Next we consider the $m$th moment of the measure $\mu$:
\begin{align*}
s_m (\mu) := \int_{\RM}x^m\, k(x\!:\!r)\,dx=
\int_{-r}^{r}  \frac{x^m\sqrt{1-r^2}}{\pi( 1-x^2)\sqrt{r^2-x^2} }\, dx.  
\end{align*}

\begin{thm}\label{thmmom}
For $m\ge 0$ we have $s_{2m+1} (\mu) = 0$ and
\begin{align*}
s_{2m} (\mu) = 1-\sqrt{1-r^2} \sum_{k=0}^{m-1}{2k\choose k}\left({r^2\over4}\right)^k.
\end{align*}
The moment generating function is equal to
\begin{equation*}
M_{\mu}(z):=\sum_{m=0}^{\infty} s_{m}(\mu)z^m
=\frac{1-r^2 z^2-z^2\sqrt{1-r^2}\sqrt{1-r^2 z^2}}{(1-z^2)(1-r^2z^2)}.
\end{equation*}
\end{thm}

We note that this result for the Hadamard walk case appeared in Konno {\it et al.} \cite{KonnoEtAl2004}. 
\smallskip

\noindent
\textit{Proof.} First we note that
\begin{equation*}
M_{\mu}(z)
=G_{\mu}(1/z)/z 
=\frac{1-r^2 z^2-z^2\sqrt{1-r^2}\sqrt{1-r^2 z^2}}{(1-z^2)(1-r^2z^2)}.
\end{equation*}
Now denoting the sequence given in theorem by $s_m$, and its generating function by $M(z)$, we have 
\begin{align*}
M(z)
&= \sum_{n=0}^\infty \left\{ 1-\sqrt{1-r^2} \sum_{k=0}^{n-1}{2k\choose k}
\left({r^2\over4}\right)^k \right\} z^{2n} 
\\
&=\frac{1}{1-z^2}-\sqrt{1-r^2}\sum_{k=0}^\infty\sum_{n=k+1}^{\infty}
{2k\choose k}\left(\frac{r^2}{4}\right)^k z^{2n} \\
&=\frac{1}{1-z^2}-\sqrt{1-r^2}
\sum_{k=0}^\infty{2k\choose k}\left(\frac{r^2}{4}\right)^k\frac{z^{2k+2}}{1-z^2} \\
&=\frac{1}{1-z^2}-\frac{z^2\sqrt{1-r^2}}{(1-z^2)\sqrt{1-4\frac{r^2z^2}{4}}} \\
&=\frac{1-r^2 z^2-z^2\sqrt{1-r^2}\sqrt{1-r^2 z^2}}{(1-z^2)(1-r^2z^2)},
\end{align*}
which is equal to $M_{\mu}(z)$.
We used the well known formula:
\begin{align*}
       \sum_{k=0}^\infty {2k\choose k}x^k={1\over\sqrt{1-4x}}.
\end{align*}

\section{Asymmetric case}
In this section, we consider an asymmetric case of the limit density $f_{\infty}(x)$.
Denote by $\mu(r,c)$ the probability measure on $\RM$ which has density $(1+ c x)\,k(x\!:\!r)$, where $r\in(0,1)$ and $c\in[-1/r,1/r]$. We note that $c=0$ leads to the symmetric case.

\begin{thm}
For the moment generating function and the Stielties
transform of $\mu(r,c)$ we have
\begin{align*}
M_{\mu(r,c)}(z)
&=\cfrac{(1-r^2 z^2)(1+cz)-(z+c)z\sqrt{1-r^2}\sqrt{1-r^2 z^2}}{(1-z^2)(1-r^2 z^2)},\\
G_{\mu(r,c)}(z)
&={(z^2-r^2)(z+c)-(1+cz)\sqrt{1-r^2}\sqrt{z^2-r^2}\over(z^2-1)(z^2-r^2)}.
\end{align*}
\end{thm}

\noindent
\textit{Proof.}
First we note that for the moments of $\mu(r,c)$ we have
\begin{align*}
s_n(\mu(r,c))
&= \int_{-r}^{r}\frac{x^n\,\sqrt{1-r^2}(1+cx)}{\pi(1-x^2)\sqrt{r^2-x^2}}\,dx\\
&=\int_{-r}^{r}\frac{x^n\,\sqrt{1-r^2}}{\pi(1-x^2)\sqrt{r^2-x^2}}\,dx
+c\int_{-r}^{r} \frac{x^{n+1}\,\sqrt{1-r^2}}{ \pi(1-x^2)\sqrt{r^2-x^2}}\,dx \\
&=s_{n}(\mu(r,0))+c\cdot s_{n+1}(\mu(r,0)).
\end{align*}
Hence we can use Theorem {\rmfamily \ref{thmmom}} to get
\begin{align*}
M_{\mu(r,c)}(z)
&=\sum_{n=0}^{\infty}s_{n}(\mu(r,c))z^n
=M_{\mu(r,0)}(z)+\frac{c}{z}\big(M_{\mu(r,0)}(z)-1\big) \\
&=\cfrac{(1-r^2 z^2)(1+cz)-(z+c)z\sqrt{1-r^2}\sqrt{1-r^2 z^2}}{(1-z^2)(1-r^2 z^2)}.
\end{align*}
From this we obtain the Stieltjes transform $G_{\mu(r,c)}(z)=M_{\mu(r,c)}(1/z)/z$ of $\mu(r,c)$.
\smallskip

Denote by $\beta_n(r,c)$, $\gamma_n(r,c)$ the Jacobi coefficients of $\mu(r,c)$, so that
\begin{align*}
G_{\mu(r,c)}(z)=\cfrac{1}{z-\beta_0(r,c)-\cfrac{\gamma_0(r,c)}{z-\beta_1(r,c)-\cfrac{\gamma_1(r,c)}{ z-\beta_2(r,c)-\cfrac{\gamma_2(r,c)}{\ddots}}}}.
\end{align*}
Since $\mu(r,-c)$ is the reflection of $\mu(r,c)$
we have $\beta_n(r,-c)=-\beta_n(r,c)$ and $\gamma_n(r,-c)=\gamma_n(r,c)$,
so we can assume that $c\ge0$.
Using combinatorial relations between moments and Jacobi coefficients (see \cite{ab,vi}) one can check that
\begin{align*}
\beta_0(r,c) &= c(1-s),\\
\gamma_0(r,c)&=(1-s)(1-c^2+c^2 s),\\
\beta_1(r,c)&={-c(1-s)(2-2c^2-s+2c^2 s)\over2(1-c^2+c^2 s)},\\
\gamma_1(r,c)&={s(1-s)(2-2c^2+c^2 s+c^2 s^2)\over4(1-c^2+c^2 s)^2},
\end{align*}
where $s:=\sqrt{1-r^2}$, and these coefficients are getting more and more complicated. It is possible however to find them in particular cases.

\begin{pro}
\begin{align*}
\beta_n (r,1)&= \left\{
\begin{array}{lc}
1-\sqrt{1-r^2}, & n=0, \\
-(1-\sqrt{1-r^2})/2, & n=1, \\
0, & n \ge 2, \\
\end{array}
\right . \nonumber \\
\gamma_n (r,1)&= \left\{
\begin{array}{lc}
\sqrt{1-r^2}(1-\sqrt{1-r^2}), & n=0, \\
r^2/4, & n \ge 1. \\
\end{array}
\right . \nonumber 
\end{align*}
\end{pro}

\noindent
\textit{Proof.}
We have 
\begin{align*}
&\cfrac{1}{z-\left(1-\sqrt{1-r^2}\right)-\cfrac{\sqrt{1-r^2}\left(1-\sqrt{1-r^2}\right)}
{z+\left(1-\sqrt{1-r^2}\right)/2-\cfrac{r^2/4}{z-\cfrac{r^2/4}{z-\cfrac{r^2/4}{\ddots}}}}}
\\
&=\cfrac{1}{z-\left(1-\sqrt{1-r^2}\right)-\cfrac{\sqrt{1-r^2}\left(1-\sqrt{1-r^2}\right)}
{z+\left(1-\sqrt{1-r^2}\right)/2-A(z)r^2/4}} 
\\
&=\cfrac{1}{z-\left(1-\sqrt{1-r^2}\right)-\cfrac{2\sqrt{1-r^2}\left(1-\sqrt{1-r^2}\right)}
{z+\left(1-\sqrt{1-r^2}\right)+\sqrt{z^2-r^2}}}
\\
&=\frac{z+\left(1-\sqrt{1-r^2}\right)+\sqrt{z^2-r^2}}
{z^2-r^2+\left(z-(1-\sqrt{1-r^2})\right)\sqrt{z^2-r^2}}
\\
&=\frac{2\left(1-\sqrt{1-r^2}\right)\left((z^2-r^2)-\sqrt{1-r^2}\sqrt{z^2-r^2}\right)}
{2\left(1-\sqrt{1-r^2}\right)(z^2-r^2)(z-1)} =G_{\mu(r,1)}(z).
\end{align*}
\smallskip

In a similar way one can prove
\begin{pro}
\begin{align*}
\beta_n (r,1/r)&= \left\{
\begin{array}{lc}
(1-\sqrt{1-r^2})/r, & n=0, \\
-(1-\sqrt{1-r^2})^2/(2r), & n=1, \\
0, & n \ge 2,
\end{array}
\right . \nonumber \\
\gamma_n (r,1/r)&= \left\{
\begin{array}{lc}
\sqrt{1-r^2}(1-\sqrt{1-r^2})^2, & n=0, \\
r^2/4, & n \ge 1.
\end{array}
\right . \nonumber 
\end{align*}
\end{pro}

\section{General case}
We consider a general case with the Jacobi coefficients $\{ \gamma_n \}_{n=0}^{\infty}$ which are given by 
\begin{align*}
p_0 = \gamma_0, \>\> p_1 = \gamma_1, \>\> \ldots, \>\> p_{n-1}=\gamma_{n-1}, \>\> p = \gamma_n = \gamma_{n+1} = \cdots.  
\end{align*}
The corresponding Stieltjes transform is denoted by $G^{(n)} (z).$ In this paper, we call the case {\it $(p_0,p_1, \ldots, p_{n-1}, p)$-case}. As we showed in Sect. 2, the symmetric case induced by the quantum walk is a special case for $n=2$, i.e., $(1 - \sqrt{1-r^2},\sqrt{1-r^2}(1- \sqrt{1-r^2})/2,r^2/2)$-case. In a similar fashion, we can obtain an explicit form of $G^{(n)} (z)$ as follows.

\begin{thm}
\begin{align*}
G^{(n)}(z)=\frac{\Pi_{n-2}(z)}{\Pi_{n-1}(z)},
\end{align*}
where $\Pi_k(z)=z\Pi_{k-1}(z)-p_{n-(k-1)}\Pi_{k-2}(z)$ with $\Pi_0(z)=z-p_{n-1}A(z), \Pi_{-1}(z)=1$. Here 
\begin{align*}
A(z)=\frac{z-\sqrt{z^{2}-4p}}{2p}.
\end{align*}
In particular, 
\begin{align*}
G^{(1)}(z) 
&= \frac{1}{2} \cdot \frac{(2p-p_0)z - p_0 \sqrt{z^2-4p}}{(p-p_0)z^2 + p_0^2},
\\
G^{(2)}(z) 
&= \frac{(2p-p_1)z+p_1 \sqrt{z^2-4p}} {(2p-p_1)z^2-2p_0p+p_1z \sqrt{z^2-4p}}, 
\\
G^{(3)}(z) 
&= \frac{(2p-p_2)z^2-2p_1p+p_2z\sqrt{z^2-4p}} {(2p-p_2)z^3+(p_0p_2-2p_0p-2p_1p)z+p_2(z^2-p_0)\sqrt{z^2-4p}}.
\end{align*}
\end{thm}
Applying the Stieltjes inverse formula, we have the following absolutely continuous part of the corresponding probability measure.
\begin{cor}  
\begin{align*}
\rho^{(1)}(x)
&=\frac{1}{2\pi} \> \frac{p_0 \sqrt{4p-x^2}} {(p-p_0)x^2+p_0^2} \> I_{(-2 \sqrt{p}, \> 2 \sqrt{p})}(x), 
\\
\rho^{(2)}(x)
&=\frac{1}{2\pi} \> \frac{p_0 p_1 \sqrt{4p-x^2}} {(p-p_1)x^4+ \{ p_0(p_1 -2p)+p_1^2\}x^2+p_0^2p} \> I_{(-2 \sqrt{p}, \> 2 \sqrt{p})}(x), 
\\
\rho^{(3)}(x)
&=\frac{1}{2\pi} \> \frac{p_0p_1p_2 \sqrt{4p-x^2}} {(p-p_2)x^6 + c_1 x^4 + c_2 x^2 +p_0^2p_1^2} \> I_{(-2 \sqrt{p}, \> 2 \sqrt{p})}(x), 
\end{align*}
where
\begin{align*}
c_1 = (p_0 + p_1)(p_2 - 2p) + (p_0 + p_2)p_2, \qquad c_2 = (p_0 + p_1)^2 p - p_0p_2 (p_0 + p_1 + 2p_2). 
\end{align*}
\end{cor}
For $n=1$ case, see Obata \cite{Obata2004}. In particular, when $p_0 = 2N$ and $p = 2N-1$, $\rho^{(1)}(x)$ is a constant multiple of the density function of a Kesten distribution. 

Finally we consider the following asymmetric case:
\begin{align*}
p_0 = \gamma_0, \>\> p_1 = \gamma_1, \>\> p = \gamma_2 = \gamma_3 = \cdots, \qquad q_0 = \beta_0, \>\> q = \beta_1 = \beta_2 = \cdots.
\end{align*}
In a similar way, we obtain
\begin{pro}
\begin{align*}
G^{(2,asym)}(z) 
&= \frac{(2p-p_1)z - q(2p-p_1)+p_1 \sqrt{(z-q)^2-4p}}{(2p-p_1)(z-q_0)(z-q)-2p_0p+p_1(z-q_0) \sqrt{(z-q)^2-4p}}, 
\\
\rho^{(2,asym)}(x)
&=\frac{1}{2\pi} \> \frac{p_0 p_1 \sqrt{4p-(x-q)^2}}{(p-p_1)(x-q_0)^2(x-q)^2+ \{ p_0(p_1 -2p)(x-q)+p_1^2 (x-q_0)\}(x-q_0)+p_0^2p} 
\\
& \qquad \qquad \times I_{(q-2 \sqrt{p}, \> q+2 \sqrt{p})}(x).
\end{align*}
\end{pro}
We note that if $q_0 = q =0$, then $G^{(2,asym)}(z)$ (resp. $\rho^{(2,asym)}(x)$) becomes $G^{(2)}(z)$ (resp. $\rho^{(2)}(x)$).
\par
\
\par\noindent
{\bf Acknowledgment.} 
Wojciech M\l otkowski
is supported by MNiSW: 1P03A~01330, by ToK: MTKD-CT-2004-013389,
by 7010 POLONIUM project:
\textit{``Non-Commutative Harmonic Analysis
with Applications to Operator Spaces, Operator Algebras and Probability",}
and by joint PAN-JSPS project:
\textit{``Noncommutative Harmonic Analysis on Discrete Structures
with Applications to Quantum Probability".}
\par
\
\par

\begin{small}
\bibliographystyle{jplain}

\end{small}

\end{document}